\documentstyle[eqsecnum,aps,prd]{revtex}
\newif\ifAMStwofonts
\begin{document}
\def\lesssim{\mathrel{\mathpalette\oversim<}}
\def\gtrsim{\mathrel{\mathpalette\oversim>}}
\def\oversim#1#2{\lower0.2ex\vbox{\baselineskip0pt\lineskip0pt
  \lineskiplimit0pt\ialign{$#1\hfil##\hfil$\crcr#2\crcr\sim\crcr}}}
\draft

\parbox{\hsize}{
\begin{flushright}
HUPD-9723, Revised Feb. 1998
\end{flushright}

\title{Boltzmann approach to the large-angle cosmic X-ray 
background fluctuations }
\author{Yasuhiro Miura, Kazuhiro Yamamoto,  
Kenta Miyauchi, and Masayasu Hosonuma} 
\address{Department of Physics, Hiroshima University,
   Kagamiyama, Higashi-Hiroshima, 739-8526~Japan}
\date{MNRAS in press}

\def\LaTeX{L\kern-.36em\raise.3ex\hbox{a}\kern-.15em
    T\kern-.1667em\lower.7ex\hbox{E}\kern-.125emX}

\maketitle
}
\vspace{10pt}

\begin{abstract}
Large-angle fluctuations in the cosmic X-ray background 
are investigated by a new formalism with a simple model of the 
X-ray sources. Our method is formulated from the Boltzmann equation 
and a simple extension of the work by Lahav et~al. to be 
applicable to a hyperbolic (open) universe.
The low multipole fluctuations due to the source clustering
are analyzed in various cosmological models in both 
numerical and analytic way.
The fluctuations strongly depend on the X-ray sources evolution 
model, as pointed out previously.
It turns out that the nearby ($z \lesssim 0.1$) sources are
the dominant contributors to the large-angle fluctuations.
If these nearby sources are removed in an observed X-ray map, 
the dipole (low multipole) moment of the fluctuation
drastically decreases.
In this case the Compton-Getting effect 
of an observer's motion can be a dominant contribution
to the dipole fluctuation.
This feature of fluctuation, relating to the matter power spectrum,
is discussed.

\noindent
{\bf keywords} ~cosmology: theory 
--- cosmic X-ray background
--- large-scale structure
\end{abstract}
\
\section{Introduction}

One of the most important problems in cosmology 
is to explain the formation of the cosmic structure.
The discovery of anisotropies in the cosmic microwave background 
(CMB) supports the hypothesis that the structure
of the Universe formed via gravitational instability from small
density fluctuations (e.g.,~White, Scott, \& Silk 1994).
It is also argued that the cosmological models with cold 
dark matter (CDM) are favorable models with which 
to explain the large-scale structure 
of the galaxy distribution and the formation of the cosmic objects
at the high-$z$ Universe 
(e.g.,~Dodelson, Gates, \& Turner 1996; Peacock \& Dodds 1994). 

The origin of the cosmic X-ray background (CXB)
radiation is also explored 
(e.g.,~Fabian \& Barcons 1992;Hayashida 1989).
The deep surveys of the 0.5-2 keV~X-ray range were carried out 
with ROSAT, and most of the background sources were resolved 
into AGN and other X-ray luminous galaxies 
(e.g., Hasinger 1997).
From the isotropy of the observation, it was concluded that a good deal of 
the unresolved fraction of the CXB has come from redshifts $z>1$ 
(e.g., Carrera, Fabian \& Barcons 1997).
The conclusion may be supported in the higher X-ray energies band with 
ASCA observations (Georgantopoulos et al. 1997).
The finite solution of the source problem will be obtained in the 
next generation of X-ray instruments (e.g., Charles \& Seward 1995).

The CXB fluctuations are an important probe 
of the large-scale structure of the universe
(~Barcons \& Fabian 1988; Carrera et~al. 1997; 
Barcons, Fabian \& Carrera 1997; Lahav, Piran \& Treyer 1997).
For example, the relation between the angular correlation function 
of the intensity fluctuations and the two-point 
spatial correlation function of galaxies $\xi(r)$ was studied. 
(De~Zotti et~al. 1990; Martin-Mirones et~al. 1991; Persic et~al. 1989;
Shafer 1983).
The cosmological models with a cosmological constant can be  
tested by the cross-correlation between the X-ray and the 
microwave background(Boughn, Crittenden \& Turok 1997).

Recently, Lahav et~al.(1997) have investigated the large-angle 
fluctuations in the CXB. They have calculated 
the expected values of the fluctuations in a statistical way. 
Treyer et~al.(1998) have compared the predicted CXB fluctuations 
with the HEAO1-A2 measurements. 
In the prediction of CXB fluctuations there are many uncertainties.
The evolution of the X-ray sources are not 
completely understood. A simple (power-law) 
source evolution model is assumed in their paper 
(Lahav et al. 1997; Treyer et al.1998).
Furthermore only a simple flat cosmological model is assumed.
The difference in the cosmological model might yield the large 
difference in amplitude of the fluctuation.   
For example, if the X-ray sources at the high-$z$ universe are the
dominant contributors to the angular fluctuations, the cosmological 
parameters, 
e.g., the curvature of the universe and the cosmological constant 
$\Lambda$, are significant factors.
It is therefore worth examining how the fluctuations depend on the 
cosmological parameters. 
We also develop a useful formalism which is applicable to a hyperbolic 
(open) universe in order to extend the work by Lahav et al. 
to various cosmological models, which is described in Section 2.
Our formalism is based on the Boltzmann approach,
which is familiar in analysis of the cosmic microwave 
anisotropies (e.g.,~Hu \& Sugiyama 1995a;1995b).
A simple model for the source distribution is introduced.
In Section 3 we solve the Boltzmann equation and obtain the expression for 
the root mean square of multipole moments of the fluctuations.
In Section 4 the fluctuations are analyzed in various cosmological models. 
Section 5 is devoted to summary and discussions.


Throughout this paper we will use the units $c=\hbar=k_B=1$;
however, we occasionally use the Planck constant $h_{\rm P}(=2\pi\hbar)$ 
to make clear the meaning of equations.
\section{Formalism}
\def\aova{{a}}
\def\falpha{{f}}
\def\malpha{{m_{(\alpha)}}}
\def\vecx{{{x^i}}}
\def\vecq{{{q^i}}}
\def\vecp{{{p^i}}}
\def\vecgamma{{{\gamma}^i}}
\def\q0{{q}}
\def\Lnu{{L_{\nu}}}
\def\bfx{{\bf x}}
\def\bfk{{\bf k}}
\def\bfq{{\bf q}}
\def\rhoX{{\rho_{\rm X}}}
\def\hplanck{{h_{\rm P}}}
\def\bX{{b_X}}
\def\j{{j}}
\def\deltac{{\delta_{\rm c}}}
\def\Vc{{V_{\rm c}}}
\def\rhoc{{\rho_{\rm c}}}
The perturbation of the Friedmann-Robertson-Walker (FRW)
space-time in the Newtonian gauge is written as 
\begin{equation}
  ds^2=-(1+2\Psi)dt^2
  +\aova^2 (1+2\Phi)\gamma_{ij}dx^idx^j ~ ,
\label{metric}
\end{equation}
where $\Psi$ is the perturbed gravitational potential, $\Phi$ is the 
curvature perturbation, $\gamma_{ij}$ is the three-metric on a space of 
constant negative curvature, and $a$ is the scale factor normalized 
to be unity at present. The scale factor $a$ is given by the Friedmann 
equation,
\begin{equation}
  \biggl({\dot a\over a}\biggr)^2=H^2_0
  \biggl( {1\over a}\Omega_0+\Omega_K+{a^2} \Omega_\Lambda\biggr)~,
\label{FriedmannEq}
\end{equation}
where $H_0$ is the Hubble parameter with $H_0=100~h~{\rm km/s/Mpc}$,
$\Omega_0$ is the density parameter, $\Omega_\Lambda(=\Lambda/3H_0^2)$ 
is defined by the cosmological constant $\Lambda$, 
and $\Omega_K(=1-\Omega_0-\Omega_\Lambda)$ describes the spatial 
curvature of the universe.
For the spatially flat models, $\Omega_K$ is equal to zero.
In equation (\ref{FriedmannEq}) the dot denotes $\eta$ differentiation, 
where $\eta$ is the conformal time defined by $a d\eta=dt$.

The propagation of photon is affected by the expansion and the
metric perturbations.
The Boltzmann equation in the perturbed FRW universe is derived in 
Appendix A. The explicit form is expressed as
\begin{equation}
  {\partial \falpha\over \partial\eta}
  +(1-\Phi+\Psi){\partial \falpha\over \partial \vecx}
  {\vecgamma}
  -\q0{\partial \falpha\over \partial \q0}
  \biggl({\dot a\over a}+{\partial\Phi\over\partial\eta}
  +{\partial\Psi\over\partial\vecx}{\gamma^i}  \biggr)
  +{\partial\falpha\over\partial\vecgamma}{d \vecgamma\over d\eta}
  =\aova S[\eta,\vecx,\q0,\vecgamma] ~, 
\label{Boltz}
\end{equation}
where $q$ is the photon energy in the locally orthonormal frame,
and $\vecgamma$ is the spatial vector 
normalized as  $\gamma_{ij}\gamma^i\gamma^j=1$ in this frame.

We now discuss the source term S in the right-hand side 
in equation (\ref{Boltz}).
The X-ray sources may be the discrete point sources. 
As we are interested in the much larger scale,
we can average the distribution of the X-ray sources 
over a certain region and regard the source function 
as a smoothed field with spatially varying emissivity.
We write the comoving volume emissivity per frequency at the time 
$\eta$ as $\j_\nu(\eta,\bfx)$ and separate it into two parts as
\begin{equation}
 \j_\nu(\eta,\bfx)=\bar \j_\nu(\eta)+ \delta\j_\nu(\eta,\bfx)~,
\end{equation}
where the bar represents a spatially averaged quantity, and $\delta j_\nu$ 
is the spatially inhomogeneous part of $j_\nu$.
In this paper, we consider the cosmological models
dominated by CDM.
The emissivity may be changed due to the presence of the inhomogeneity
of the CDM.
We expect that the value $j_\nu$ should increase as the CDM density
fluctuation $\deltac$ ($= (\rhoc-\bar\rhoc)/ \bar\rhoc$) becomes large.
Therefore, the emissivity in the inhomogeneous universe is expressed as
\begin{equation}
 \j_\nu(\eta,\bfx)=\bar \j_\nu(\eta) \Bigl(1+b_X\deltac\Bigr)~,
\end{equation}
where $\bX$ is the bias factor to relate the large-scale 
distribution of the CDM with that of the X-ray sources.
\footnote{
It may be unrealistic to assume the epoch-independent biasing.
The epoch-dependent biasing is assumed in the recent paper
by Treyer et~al.(1998) by setting $b_X(z)=b_X(0)+z[b_X(0)-1]$~.
As we describe in section 4 and Appendix D, however,
the nearby sources ($z \lesssim 0.1$) are the dominant contributors 
to the large-angle fluctuations $(l\lesssim 10)$.
In this case the amplitude of large-angle fluctuations
will roughly scale in proportion to $b_X(0)$ unless
$b_X(0)$ becomes very large.
}
In this paper we assume 
\begin{equation}
  \bar\j_\nu(\eta)=j_0(\nu)E(z)~,
\label{simpleform}
\end{equation}
with $E(z)=(1+z)^p$ for redshift $z_{\rm min}\leq z\leq z_{\rm max}$, 
and $E(z)=0$ for other period.  Note that $p$ is constant and 
$p=0$ is the case of no evolution of the X-ray sources. 
We also assume the power-law 
form of the energy spectrum, $j_0(\nu)\propto \nu^{-\alpha}$ 
with $\alpha=0.4$.
Here we set $\alpha$ to be the observed CXB spectral index
in (3-20)~keV range (Comastri et al. 1995), however, a reasonable change 
in $\alpha$ will not affect our conclusions at a significant level.

The photon number emitted during the proper time 
$t\sim t+dt$, in the frequency range $\nu\sim\nu+d\nu$,
in the solid angle $d\Omega$, from the volume $\bfx\sim \bfx+d\bfx$, 
is written as
\footnote{
The fluctuation of the metric might have to be taken into 
account when formulating of the source term. As we will see 
in the next section (also Lahav et~al. 1997), however,
the contribution from the metric fluctuations to the 
large-angle fluctuations in the CXB is small.
Therefore we have omitted the metric fluctuations in the source term.
}
\begin{equation}
  {\j_\nu\over \hplanck \nu}\Bigl(1+(3+\alpha)\gamma_i V^i_X \Bigr)
  ~d\nu {d\Omega\over 4\pi} d^3\bfx dt~,
\label{sourceA}
\end{equation}
where we assumed that this volume element has the peculiar velocity $V_X$,
and note that the Doppler effect is taken into account.

From the definition of the source term $S$ in the Boltzmann equation,
the photon number emitted during $t\sim t+dt$, in the momentum 
range $\bfq\sim\bfq+d\bfq$, from the volume $\bfx\sim \bfx+d\bfx$, 
is $Sg{d^3\bfq}{a}^3 d^3\bfx dt/\hplanck^3$, where 
$g(=2)$ is the photon statistical weight.
Using the relation $q(=|{\bf q}|)=\hplanck\nu$, 
the photon number emitted during $t\sim t+dt$, 
in the frequency range $\nu\sim\nu+d\nu$,
from the volume $\bfx\sim \bfx+d\bfx$, in the solid angle $d\Omega$, 
is written as $g S \nu^2d\nu {d\Omega} a^3d^3\bfx dt$. 
Equating this expression and equation (\ref{sourceA}), we find
\begin{equation}
  S={1\over g 8\pi^2}{1\over a^3} {\j_\nu \over \nu^3}
  \Bigl(1+(3+\alpha)\gamma_i V^i_X \Bigr),
\label{sourceD}
\end{equation}
where we used the relation $\hplanck=2\pi$ in our unit.
The peculiar velocity of the X-ray sources $V_X$ does
not necessarily agree with that of the CDM, $\Vc$.
We introduce another bias factor $b_V$ for velocity, where 
$\Vc$ is the CDM velocity field.

We eventually get the source term as 
\begin{eqnarray}
  S={1\over g8\pi^2}{1\over a^3} {\bar\j_\nu \over \nu^3}
  \Bigl(1+b_X \deltac(\eta,\bfx)
  +(3+\alpha) b_V \gamma_i \Vc^i(\eta,\bfx)\Bigr).
\end{eqnarray}


\section{Solution of the Boltzmann Equation} 
\def\Delt{{\vartheta}}
\def\Thetal{{\Theta_l}}
\def\Ylm{{Y_{lm}}}
\def\tk{{\tilde k}}
In order to solve the above Boltzmann equation (3), we use the linear 
perturbative expansion method with respect to the inhomogenity (such as
metric perturbations). If the universe is completely homogeneous, 
the distribution function $f^{(0)}$ and the source term $S^{(0)}$ can be 
described by $\eta$ and $\q0$. With the inhomogenity, they are modified as
\begin{eqnarray}
  &&f=f^{(0)}(\eta,\q0)+f^{(1)}(\eta,\vecx,\q0,\vecgamma)~,
\\
  &&S=S^{(0)}(\eta,\q0)+S^{(1)}(\eta,\vecx,\q0,\vecgamma)~,
\end{eqnarray}
with,
\begin{eqnarray}
  &&S^{(0)}(\eta,\q0)={1\over g8\pi^2}{1\over a^3} 
  {\bar\j_\nu \over \nu^3}~,
\\
  &&S^{(1)}(\eta,\vecx,\q0,\vecgamma)=
  S^{(0)}(\eta,\q0)\Bigl(b_X \deltac(\eta,\vecx)
  +(3+\alpha) b_V \gamma_i \Vc^i(\eta,\vecx)\Bigr)~.
\end{eqnarray}
The Boltzmann equations of the zeroth and first order are given by
\begin{equation}
  {\partial f^{(0)}\over\partial\eta}-
  \q0{\partial f^{(0)}\over\partial \q0}{\dot a\over a}
  =\aova S^{(0)} ~ ,
\label{homoA}
\end{equation}
and
\begin{eqnarray}
  {\partial f^{(1)}\over\partial\eta}
  +{\partial f^{(1)}\over\partial x^i} \gamma^i
  -\q0{\partial f^{(1)}\over\partial \q0}{\dot a\over a}
  -\q0{\partial f^{(0)}\over\partial \q0}            
  \biggl({\partial\Phi\over\partial\eta}
  +{\partial \Psi\over\partial x^i}\gamma^i \biggr)
  +{\partial f^{(1)}\over \partial \gamma^i} {d \gamma^i\over d\eta}
  =\aova S^{(1)} ~ .
\label{pertA}
\end{eqnarray}
Note that when neglecting the metric perturbations, 
equations (\ref{homoA}) and (\ref{pertA}) are equivalent 
to the original Boltzmann equation, and give the exact solution. 

Introducing the comoving momentum by $Q=aq$,
the above equations reduce to
\begin{equation}
  {\partial f^{(0)}\over\partial\eta}
  =\aova S^{(0)}(\eta,Q/a) ~ ,
\label{zeroth}
\end{equation}
and
\begin{eqnarray}
  {\partial f^{(1)} \over\partial\eta}
  +\gamma^i{\partial f^{(1)}\over\partial x^i}  
  -Q{\partial f^{(0)} \over\partial Q} 
  \biggl({\partial\Phi\over\partial\eta}
  +{\partial \Psi\over\partial x^i}\gamma^i\biggr)
  +{\partial f^{(1)}\over \partial \gamma^i} 
  {d\gamma^i\over d\eta}
  =\aova S^{(1)}(\eta,\vecx,Q/a,\vecgamma)~,
\label{pertB}
\end{eqnarray}
where $f^{(0)}$ and $f^{(1)}$ must be regarded as
the functions $f^{(0)}(\eta,Q)$ and 
$f^{(1)}(\eta,\vecx,Q,\vecgamma)$, respectively.

Let us first consider the zeroth order equation (\ref{zeroth}),
which is formally integrated as
\begin{eqnarray}
  f^{(0)}(\eta,Q)&=&\int^\eta_0 d\eta' 
  ~{a'} S^{(0)}(\eta',Q/a')
\nonumber
\\
  &=&{\pi\over g} Q^{-3} \int_0^\eta d\eta'a'
  \bar j_{Q/h_{\rm P}a'}(\eta') ~,
\label{zerothA}
\end{eqnarray}
where we used  
  $S^{(0)}(Q/a)=(\pi/ g Q^{3}) 
  \bar j_{Q/h_{\rm P}a}(\eta)$,
  and defined $a'=a(\eta')$. 
Since we have assumed the power-law form, 
$\bar j_\nu(\eta)= j_0(\nu)(1+z)^p$,  
for $z_{\rm min}\leq z\leq z_{\rm max}$, (see equation~(\ref{simpleform})), 
the specific intensity, $I^{(0)}_\nu=g h_{\rm P} \nu^3 f^{(0)}$ 
(e.g.,~Shu 1991) is written as 
\begin{equation}
  I^{(0)}_\nu(\eta_0)
  ={\j_{0}(\nu)\over 4\pi H_0} \int_{z_{\rm min}}^{z_{\rm max}}
   {dz(1+z)^{p-2-\alpha}\over 
  \sqrt{\Omega_0(1+z)+\Omega_K+\Omega_\Lambda/(1+z)^2}}~,
\end{equation}
where we used the relation between the conformal 
time and the redshift (\ref{FriedmannEq}).

We next consider equation (\ref{pertB}). 
It is convenient to introduce a new variable 
$\Theta(\eta,\vecx,Q,\vecgamma)$ as 
\begin{equation}
\Theta(\eta,\vecx,Q,\vecgamma)
={f^{(1)}\over f^{(0)}}={\Delta I_\nu \over I_\nu^{(0)}}~,
\end{equation} 
where we used the general relation $I_\nu=g h_{\rm P} \nu^3 f$ for  
the second equality. By using $\Theta(\eta,\vecx,Q,\vecgamma)$, 
equation (\ref{pertB}) can be written as
%
\begin{eqnarray}
  &&{\partial\Theta\over\partial\eta}+\xi(\eta) 
  \Theta+{\partial\Theta\over\partial x^i} 
  \gamma^i+ (\alpha+3)\biggl({\partial\Psi\over\partial x^i} 
  \gamma^i+{\partial\Phi\over\partial\eta} \biggr) 
  +{\partial \Theta\over \partial \gamma^i}{d \gamma^i\over d\eta}
\nonumber
\\
  &&\hspace{5cm}
  =\xi(\eta)
  \Bigl(b_X \deltac(\bfx)+b_V (3+\alpha)\gamma_i
  \Vc^i(\bfx)\Bigr)~,
\label{pertC}
\end{eqnarray}
where we used the relation 
$Q\partial f^{(0)}/\partial Q=-(\alpha +3)f^{(0)}$ and equation 
(\ref{zeroth}), and $\xi$ is defined by
$\xi(\eta)=\bigl(\partial\dot f^{(0)}/\partial Q\bigr)
  / \bigl(\partial f^{(0)}/\partial Q\bigr)={\dot f^{(0)} / f^{(0)}}
={\dot \varepsilon(\eta) / \varepsilon(\eta)}$ with
\begin{equation}
 \varepsilon(\eta)
  =\int_0^\eta d\eta'~ a' j_{Q/h_{\rm P}a'}(\eta')
  \propto \int_0^\eta d\eta' E(\eta') a'{}^{1+\alpha}~.
\label{defepsilon}
\end{equation}
Note that the variable $Q$ is separated in equation (\ref{pertC}) 
and that $\Theta$ can be regard as the function $\Theta(\eta,\bfx,\gamma^i)$.
This is the result of the simple assumption of the power-law energy spectrum
(\ref{simpleform}).

Equation (\ref{pertC}) can be written as
\begin{eqnarray}
  &&{d\over d\eta}\biggl[
  \varepsilon\Bigl(\Theta(\eta,\bfx(\eta),\gamma(\eta))
  +(\alpha+3)\Psi(\eta,\bfx(\eta))\Bigr)\biggr]
\nonumber
\\
  &&\hspace{2cm}
  =2(\alpha+3)\varepsilon^{1/2}{\partial \over \partial \eta}
  \Bigl(\varepsilon^{1/2}\Psi(\eta,\bfx(\eta))\Bigr) 
  +\dot\varepsilon \Bigl( b_X \deltac
(\eta,\bfx)+b_V(3+\alpha)\gamma_i 
\Vc^i\Bigr)~,
\label{pertE}
\end{eqnarray}
where we used $\Phi=-\Psi$ from the linear perturbation theory. 
This can be easily integrated as
\begin{eqnarray}
  &&\Theta(\eta_0,\bfx(\eta_0),\gamma)
  +(\alpha+3)\Psi(\eta_0,\bfx(\eta_0))
\nonumber
\\
  &&\hspace{0.5cm}
  ={1\over \varepsilon(\eta_0)}\int_0^{\eta_0}d\eta
   ~\biggl[2(\alpha+3)\varepsilon^{1/2}{\partial \over \partial\eta}
\Bigl(\varepsilon^{1/2}\Psi(\eta,\bfx)\Bigr) +\dot\varepsilon 
\Bigl( b_X \deltac (\eta,\bfx)+b_V(3+\alpha)\gamma_i \Vc^i\Bigr)\biggr]~,
\label{integrate}
\end{eqnarray}
where we used $\varepsilon(0)=0$. 

We rewrite this solution by using the multipole expansion 
method. The formalism for the CMB
anisotropies is useful (e.g., Hu \& Sugiyama 1995b).
Since $\Theta(\eta,\bfx,\gamma^i)$ follows the linear equation,
we can expand it as a sum of modes labeled by $({\bf k},l)$,
\begin{equation}
  \Theta(\eta,\bfx,\gamma^i)=\sum_{\bf k} \sum_{l} \Thetal(\eta,k)
  G_l(\bfx,\gamma^i)~,
\label{multipoledec}
\end{equation}
where $G_l$ is a certain mode function. 
The explicit form is written in Appendix B.
The solution for $\Theta_l(\eta_0,k)$ can be found from equation 
(\ref{integrate}):
\begin{eqnarray}
 &&\hspace{-0.7cm}
{\Theta_0(\eta_0,k)}=-(\alpha+3)\Psi(\eta_0,k)
+{1\over \varepsilon(\eta_0)}\int_0^{\eta_0}
  d\eta \biggl[ 2(\alpha+3)\varepsilon^{1/2}
  {\partial\over\partial\eta}\Bigl(\varepsilon^{1/2}\Psi(\eta,k)\Bigr)
   X_\omega^0(\Delta\eta)
\nonumber
\\
  &&\hspace{2cm}
  +\dot\varepsilon b_X \deltac(\eta,k) X_\omega^0(\Delta\eta)
  -\dot\varepsilon b_V(3+\alpha) \Vc(\eta,k)X_\omega^{1}(\Delta\eta)
  \biggr]~,
\label{monop}
\end{eqnarray}
and
\begin{eqnarray}
 &&\hspace{-0.7cm}
{\Theta_l(\eta_0,k)\over 2l+1}={1\over \varepsilon(\eta_0)}\int_0^{\eta_0}
  d\eta \biggl[ 2(\alpha+3)\varepsilon^{1/2}
  {\partial\over\partial\eta}\Bigl(\varepsilon^{1/2}\Psi(\eta,k)\Bigr)
   X_\omega^l(\Delta\eta)
  +\dot\varepsilon b_X \deltac(\eta,k) X_\omega^l(\Delta\eta)
\nonumber
\\
  &&\hspace{0cm}
  +\dot\varepsilon b_V(3+\alpha) \Vc(\eta,k)
  \biggl({l\over 2l+1} X_\omega^{l-1}(\Delta\eta)
  -{l+1\over 2l+1}\Bigl(1-{l(l+2)K\over k^2}
  \Bigr)X_\omega^{l+1}(\Delta\eta)
  \biggr)\biggr]~\hspace{0.3cm}(l\ge 1)~,
\label{solut}
\end{eqnarray}
with $\Delta\eta=\sqrt{-K}(\eta_0-\eta)$ and 
\begin{equation}
  X_\omega^{l}(\chi)=\left({\pi(\omega^2+1)^{l} \over 2\sinh\chi}\right)^{1/2}
  P^{-l-1/2}_{i\omega-1/2}(\cosh\chi)~,
\label{defofX}
\end{equation}
where $P_{\nu}^\mu(x)$ is the Legendre function, and 
$\omega$ is defined as $\omega^2=k^2/(-K)-1$ with the eigen-value
$k^2$ of scalar harmonics and the spatial curvature parameter $K
=-H_0^2(1-\Omega_0-\Omega_\Lambda)$.
We can check that equations (\ref{monop}) and (\ref{solut}) satisfy 
the perturbation equations for the multipole moments derived 
in Appendix B.

In the limit of flat universe, i.e. $K \to 0$, 
the radial function $X_\omega^{l}(\chi)$ reduces to the 
spherical Bessel function, and equation (\ref{solut}) reduces to
\begin{eqnarray}
  &&\hspace{-0.7cm}{\Theta_l(\eta_0,k)\over 2l+1}
  ={1\over \varepsilon(\eta_0)}\int_0^{\eta_0}
  d\eta \biggl[ 2(\alpha+3)\varepsilon^{1/2}
  {\partial\over\partial\eta}\Bigl(\varepsilon^{1/2}\Psi(\eta,k)\Bigr) 
  j_{l}(k r_c)
  +\dot\varepsilon b_X \deltac(\eta,k)j_{l}(k r_c)
\nonumber
\\
  &&\hspace{2cm}
  +\dot\varepsilon b_V (3+\alpha) \Vc(\eta,k)
  \Bigl({l\over 2l+1}j_{l-1}(k r_c)
  -{l+1\over 2l+1}j_{l+1}(k r_c)\Bigr)\biggr]
~\hspace{0.3cm}~(l\ge1)~,
\label{solutflat}
\end{eqnarray}
where $r_c$ is the path of a photon, defined as $r_c=\eta_0-\eta$.


As we are interested in the statistics of the angular fluctuations
at a point of an observer $\bfx_0=\bfx(\eta_0)$, we consider the
two point angular correlation function 
$\bigl<\Theta(\eta_0,\bfx_0,\vecgamma)
\Theta(\eta_0,\bfx_0,\vecgamma')\bigr>$.
Expanding the fluctuation $\Theta$ at the point $\bfx_0$ by the 
spherical harmonics,
\begin{equation}
  \Theta(\eta_0,\bfx_0,\vecgamma)=\sum_{lm} a_{lm} 
  \Ylm(\Omega_\gamma),
\end{equation}
we can express the ensemble average of the 
two point angular correlation function as
\begin{equation}
  \bigl<\Theta(\eta_0,\bfx_0,\vecgamma)
  \Theta(\eta_0,\bfx_0,\vecgamma')\bigr>
  ={1\over 4\pi}\sum_{l=0} (2l+1)C_lP_l(\cos\theta)~,
\end{equation}
where we defined $C_l=\bigl<a_{lm}^2\bigr>$,  and $\cos\theta
=\vecgamma\cdot\vecgamma'$.
Here the angular power spectrum $C_l$ is given by
(e.g., Hu \& Sugiyama 1995b)
\begin{equation}
  C_l={2\over \pi}\int_0^\infty {d\tk}\tk^2
  {M_l} \biggl<\biggl\vert {\Theta_l(\eta_0,k)\over 2l+1}
\biggr\vert^2  \biggr>~,
\label{Cl}
\end{equation}
where $M_l=(\tk^2-K)\cdots (\tk^2-l^2K)/(\tk^2-K)^l$ and $\tk^2=k^2+K$.

As we see in equations (\ref{Cl}) with (\ref{solut}) or (\ref{solutflat}),
the angular power spectrum $C_l$ consists of three terms, 
which are due to the source clustering $\delta_c$,
the bulk motion of the sources $V_c$, 
and the gravitational potential fluctuations $\Psi$. 
As we will see in the below, the term proportional to 
$\deltac$ in the right-hand side  
in equations (\ref{solut}) and (\ref{solutflat}) are the 
dominant contributors to the fluctuations in the CXB (Lahav et~al 1997).
Note that the effect of the motion of an observer, 
i.e. Compton-Getting effect (hereafter C-G effect)
is not taken into account in this solution.
We integrate the above equation numerically.
The results are described in the next section.

%
%

\section{Fluctuations}
\def\zmin{{z_{\rm min}}}
\def\zmax{{z_{\rm max}}}
In this section we present our results.
We here consider three cosmological models, 
a standard CDM model (SCDM), a CDM model with a cosmological 
constant $\Lambda$ ($\Lambda$CDM), and  an open CDM model (OCDM). 
As for the initial density power spectrum, we use the 
Harrison-Zeldovich power spectrum for SCDM and $\Lambda$CDM 
with the COBE normalization (Bunn \& White 1997). 
In the OCDM model, we use the spectrum predicted in a simple 
open inflation model (Yamamoto \& Bunn 1996; White \& Silk 1996).
The spectrum is almost the same as the Harrison-Zeldovich one
at the scale of the large-scale structure, 
so that our choice of the spectrum does not 
essentially alter the results. 
In Appendix C an useful formula for the evolution of the CDM 
perturbation is summarized.


We have quite a lot of freedom in the parametrization of the 
X-ray background model. 
The parameters are classified into two kinds.
One is the cosmological parameters and the other is the model 
parameters of the X-ray sources. 
We present how the results depend on the two kinds of parameters. 

\subsection{dependence on the cosmological models}

Before discussing the dependence on the cosmological parameters,
we mention the physical contents of the fluctuations.
Figure~1 shows the angular power spectrum $C_l$ due to 
the source clustering, bulk motion, and gravitational potential
fluctuations. 
It is apparent that the source clustering term
is the dominant contributor to the large-angle fluctuations.
We have checked that the other terms due to the gravitational potential 
and the bulk motion become important only when $z_{\rm min} >0.1$. 
We therefore take only the source clustering into account hereafter, 
focusing on the case $z_{\rm min}\lesssim0.1$.


Now let us discuss the dependence of the fluctuations on the 
cosmological parameters. Fig.~2 shows the dependence of 
the dipole anisotropy $C^{1/2}_{l=1}$ on $\Omega_0$.
As we are not interested in the spectrum shape of the angular 
fluctuations here, we focus on the amplitude of $C_{l=1}$.
As expected, $\Lambda$CDM and OCDM agree with SCDM in the limit 
of $\Omega_0 \to 1$.
It is shown that the dipole anisotropy is not sensitive 
to the cosmological parameters except for the case of extreme low 
value of $\Omega_0$ (Lahav et al. 1997). 
As we see in Appendix D, this feature of the 
dipole anisotropy is attributed to the difference of the shape of 
the power spectrum. Thus $\Omega_0$ dependence can be changed 
by the normalization scheme of the matter power spectrum. 
The other point to note is that the amplitude of the fluctuations 
strongly depends on the source parameter $z_{\rm min}$. 

\subsection{dependence on the X-ray sources models}

We have three parameters, $z_{\rm min}$, $z_{\rm max}$ and $p$, 
as the X-ray sources model. 
The evolution of the X-ray sources is governed by $p$ and $z_{\rm max}$.
From now on, we set the cosmological parameters $(\Omega_0=1,~h=0.5)$
for SCDM, $(\Omega_0=0.3,~\Omega_\Lambda=0.7,~h=0.7)$ for $\Lambda$CDM, 
and  $(\Omega_0=0.3,~\Omega_\Lambda=0,~h=0.7)$ for OCDM.
In Appendix D, we described an analytic calculation for the
angular power spectrum $C_l$.
The analytic calculation is limited only to
the case of $\Omega_0=1$. However, this is instructive to understand 
the behaviour of the angular power spectrum.
Fig.~3 shows the dependence of $C_{l=1}^{1/2}$ on the evolution 
parameter $p$. 
The dipole anisotropy drastically drops with $p$. 
The fluctuation is roughly determined by dividing 
the anisotropic X-ray flux $\Delta I_\nu$ by the isotropic 
one $I^{(0)}_\nu$. 
As shown in Appendix D, the dependence on $p$ comes 
from the isotropic component $I^{(0)}_\nu$ of the CXB.
The increase of $p$, i.e, the increase of
the redshift evolution of sources leads to the relative increase of 
$I^{(0)}_\nu$ compared with $\Delta I_\nu$.
Thus the fluctuation decreases with $p$.
Fig.~4 shows the dependence of the dipole anisotropy on $z_{\rm max}$.
The amplitude of the fluctuation also decreases with $z_{\rm max}$.
The reason is the same as $p$, but the dependence 
on $z_{\rm max}$ is weaker than that on $p$.  
These features are common to the higher  multipoles (Lahav et~al. 1997).

Finally, we focus on $z_{\rm min}$ which was fixed on zero 
in the paper by Lahav et~al. (1997). If the nearby bright X-ray 
sources were removed from the observational 
data, it would not necessarily be fixed at $z_{\rm min}=0$. Fig.~5 
shows the dependence of the dipole anisotropy on  $z_{\rm min}$.
It is apparent that the dipole amplitude is
sensitive to this parameter especially for $z_{\rm min} \lesssim 0.1$.
The amplitude of the dipole moment rapidly drops with 
$z_{\rm min}$ for $z_{\rm min}\lesssim 0.1$. In contrast, 
it becomes almost constant for $z_{\rm min}\gtrsim 0.1$. 
This strong dependence on $z_{\rm min}$ is related with the shape 
of the matter power spectrum $P(k)$ as discussed in Appendix D.
The choice of $z_{\rm min}$ changes the shape of the 
angular power spectrum of the fluctuations (see Fig.~1).
This means that the sources distributed relatively close to the our 
Galaxy are the dominant contributors to the fluctuations in the CXB for 
the low multipoles.
The peak of the matter power spectrum is located at 
$\lambda\sim 100 h^{-1}{\rm Mpc}$, which corresponds to $z\sim 0.03$. 
The shape of the matter power spectrum is the reason
why the sources at the low redshift $z\lesssim 0.1$ are the
dominant contributors to the large-angle fluctuations in the CXB.

\section{Summary and Discussions}
In this paper, we have investigated the large-angle 
fluctuations in the CXB due to the X-ray source clustering. 
We have developed the formalism to describe 
the CXB fluctuations using Boltzmann equation under a simple 
model of the X-ray sources. Our formalism is a simple extension 
of that by Lahav et~al. (1997) to be applicable to an 
universe with hyperbolic geometry. 
The dependence of the fluctuations on the model parameters 
has been examined in various cosmological models.
The fluctuation does not strongly depend on the 
cosmological parameters. It is quite sensitive to the 
parameters for the X-ray sources (Lahav et~al. 1997). 
The fluctuation is determined by the ratio
of anisotropic X-ray flux to the isotropic one, i.e.,
$\sqrt{C_l}\sim \Delta I_\nu/ I^{(0)}_\nu$. 
$\Delta I_\nu$ is essentially determined by the nearby 
sources at low redshift for the low multipole moment, 
while $I^{(0)}$ is done by the high-$z$ sources.
The large redshift evolution of the X-ray emissivity (e.g., $p=3$, 
and $z_{\rm max}$ is large) makes the flux from the far sources 
large and, as a result, the amplitude of fluctuation becomes small. 

We have also pointed out the importance of the source distribution
parameter $z_{\rm min}$.
The low multipole anisotropies are sensitive to it. 
This parameter is closely related to the reconstruction 
of the X-ray map by removing nearby bright sources. 
The dipole moment with $z_{\rm min}=0.1$ 
is smaller by order of magnitude compared with the case 
$z_{\rm min}=0$.
If the sources sufficiently close to the our Galaxy 
$z~\sim 0.1$ are removed,
it may be expected that the effect of the peculiar 
motion of the observer (Compton-Getting (C-G) effect) 
is dominant in the CXB dipole. 

Let us roughly compare the effects 
which contribute to the dipole anisotropies except for the source 
clustering effect.  
The observer's peculiar motion relative to the CMB was measured 
by using the COBE four-year data (Lineweaver et al. 1996), 
which gave the peculiar velocity $V_{\rm obs}=368.9 \pm 2.5 ~{\rm km/s}$ 
in the direction ($l=264^{\circ}; b=48^{\circ}$).
Assuming we have a similar motion relative to the CXB, the expected C-G
dipole is estimated as $\sqrt{C_{l=1}^{{\rm CG}}}\simeq 5.0 \times 10^{-3}$. 

On the other hand, it is well known that the shot noise fluctuation arises 
from the discreteness of the sources. Since this fluctuation is originated
from the Poisson fluctuation of the discrete sources, the spectrum 
is white noise. The shot noise fluctuations in the CXB have been
investigated by Lahav et al. (1997).
Following their result, the amplitude of the fluctuation
is estimated $\sqrt{C_l^{\rm SN}}\simeq 1.2 \times f_{\rm m}^{1/4}$,
where $f_{\rm m}$ is a flux cut-off of bright removed sources in 
unit of ${\rm erg~s^{-1}~cm^{-2}}$. If the flux cut-off level becomes lower, 
the amplitude of shot noise decreases. Thus the shot noise fluctuation 
depends on the flux cut-off in observational data and is important 
when comparing a theoretical model with the observed X-ray map 
(Treyer et al.~1998).
The flux cut-off was $\simeq 3\times 10^{-11}~{\rm erg~s^{-1}~cm^{-2}}$ in
the HEAO-1, and the amplitude of shot noise is   
estimated as $\sqrt{C_{l=1}^{{\rm SN}}} \simeq 2.8 \times 10^{-3}$.
Note also that the cut-off level is relevant to
the parameter $z_{\rm min}$ in the sense of bright source removability.

The dipole owing to the source clustering is shown in 
Fig.5 ($p=3, z_{max}=3$).
In the case of $z_{\rm min}=0$, the dipole anisotropy due to 
the clustering effect is comparable      
to the C-G effect. However, if $z_{\rm min}$ is $O(0.1)$, the C-G dipole 
becomes well above the dipole due to the source clustering.
If the flux cut-off becomes lower by the improvement of observation, 
the shot noise and the clustering effect could be sufficiently 
smaller than C-G effect to yield a good chance of measuring the 
C-G dipole in the CXB.
This investigation also suggests that the careful treatment is required 
when comparing the observational map with the theoretical prediction.
When the redshifts of the sources can not be determined, 
the subtraction of nearby X-ray sources from the
observed map may contain a delicate problem because the nearby
faint sources may contribute to the fluctuations sensitively. 
These problems are left as future problems.

\section*{Acknowledgments}
We thank Prof. Y.Kojima for careful reading of the manuscript and 
useful comments and discussions.
We acknowledge helpful conversations on the subject of this paper
with Dr. Y.Suto, S.Sasaki, N.Sugiyama, R.Nishi, K.Hayashida, T.Tsuru, 
T.Ohsugi, and T.Yamada.
This work is supported by the Grants-in-Aid for 
Scientific Research of Ministry of Education, Science
and Culture of Japan (No.~09740203).

\newpage

\vspace{5mm}
\newpage
\appendix
\section{Boltzmann equation}
We write the Boltzmann equation for the
distribution function $\falpha(t,\vecx,\q0,\vecgamma)$ 
of photons as 
\begin{equation}
  {\partial\falpha\over\partial t}
  +{\partial\falpha\over\partial\vecx}{d\vecx\over dt}
  +{\partial\falpha\over\partial \q0}{d \q0\over dt}
  +{\partial\falpha\over\partial\vecgamma}{d\vecgamma\over dt}
  =S(t,\vecx,\q0,\vecgamma) ~ ,
\label{BoltzmannC}
\end{equation}
with the source term $S$.
Here we defined the energy in the locally orthonormal frame as
$\q0=\bigl\vert{\vecq}\bigr\vert$, 
thus $(\q0,\vecq)$ forms the 4-momentum of the photon in this 
frame.
Since the Boltzmann equation is written in terms of the momentum $q$
measured by an observer in the cosmological rest frame, we must 
rewrite the terms $d x^i/ d t$ and $d q/ d t$ in terms of $x^i$ 
and $q$ in order to solve this equation.
These relations are given from the geodesic equation of a photon. 
However, the geodesic 
equation is commonly written in terms of the 4-momentum $p^\mu$
in the frame (\ref{metric}), where $p^\mu$ is defined by 
$p^\mu=dx^\mu/d\lambda$ with the affine parameter $\lambda$.
The 4-momentum $(\q0,\vecq)$ in the rest frame
is related to the 4-momentum $p^\mu$, as follows:
\begin{eqnarray}
  &&\q0
  =(1+\Psi)p^0 ~ ,
\label{relqzero}
\\
  &&q^i
  =\aova (1+\Phi)p^i ~.
\label{relvecq}
\end{eqnarray}
The spatial vector 
$\vecgamma$ is defined as $\vecgamma=\vecq/q$ and $\vecgamma$ 
satisfies $\gamma_{ij}\gamma^i\gamma^j=1$.

Equations  (\ref{relqzero}) and (\ref{relvecq})  give the following
relations, up to the first order of $\Psi$ and $\Phi$: 
\begin{equation}
  {d\vecx\over dt}={\vecp\over p^0}
  ={1\over a}(1+\Psi-\Phi)\vecgamma ~ ,
\label{dxdt}
\end{equation}
and
\begin{equation}
  {d\q0 \over dt}=
  \biggl({\partial \Psi\over\partial t}
  +{\partial \Psi\over\partial x^j}{d x^j\over dt}\biggr)p^0
  +(1+\Psi){d p^0\over dt}.
\label{dqovdt}
\end{equation}

Meanwhile the geodesic equation in the first  order of the
perturbation is
\begin{equation}
  {d p^0\over dt}=
  \q0\biggl(
  -{1\over a}{d a\over dt}-\dot\Psi
  -{2\over a}{\partial\Psi\over\partial\vecx}
  \gamma^i-\dot\Phi+{1\over a}{d a\over dt}\Psi
  \biggr),
\end{equation}
Inserting this into (\ref{dqovdt}), we have
\begin{equation}
  {d\q0 \over dt}=-\q0\biggl(
  {1\over a}{d a\over dt}+\dot\Phi
  +{1\over a}{\partial\Psi\over\partial\vecx}{\gamma^j}  \biggr).
\label{dqovdtb}
\end{equation}
Thus we can write down the left-hand side of the Boltzmann equation
(\ref{BoltzmannC})
using equations (\ref{dxdt}) and (\ref{dqovdtb}). 
If we employ the conformal time defined by 
$a d\eta=dt$ instead of the proper time, it becomes
\begin{eqnarray}
  &&{\partial \falpha\over \partial\eta}
  +(1-\Phi+\Psi){\partial \falpha\over \partial \vecx}
  {\vecgamma}
  -\q0{\partial \falpha\over \partial \q0}
  \biggl({\dot a\over a}+{\partial\Phi\over\partial\eta}
  +{\partial\Psi\over\partial\vecx}{\gamma^j}  \biggr)
  +{\partial\falpha\over\partial\vecgamma}{d\vecgamma\over d\eta}
  =\aova S(t,\vecx,\q0,\vecgamma) ~.
\label{Boltzappendix}
\end{eqnarray}

\section{mathematical formulae in an open universe}
In this appendix, we summarize the mathematical formulae which are
needed in section 3. The results of this section
are based on the previous work (e.g., Gouda, Sugiyama, \& Sasaki 1991; 
Hu \& Sugiyama 1995b; Wilson 1983). 

We first consider the scalar harmonics $Q_\bfk(\bfx)$ on a hyperbolic
universe. It follows $Q_\bfk{}_{|i}^{~~|i}=-k^2 Q_\bfk$, 
where $|i$ denotes the covariant derivative in the hyperbolic space
with the line element
\begin{equation}
  \gamma_{ij}dx^idx^j={1\over -K} (d\chi^2+\sinh^2\chi d\Omega_{(2)}^2),
\end{equation}
where $K=-H_0^2(1-\Omega_0-\Omega_\Lambda)$ 
is the spatial curvature parameter and 
$d\Omega_{(2)}^2$ is the line element on an unit sphere.
Since we assumed that the emitted photons do not scatter, 
the trajectory of the photon is the free streaming.
Therefore, the photon geodesics are radial, we may use the radial 
scalar harmonics to describe the free streaming behaviour.
It can be expressed as $Q_\bfk=X^l_\omega(\chi)Y_{lm}(\Omega)$ using
the radial function $X^l_\omega$ in equation (\ref{defofX}) and the spherical 
harmonics  $Y_{lm}$.

The function $G_l$ for the multipole decomposition in 
equation (\ref{multipoledec}) is defined
\begin{equation}
  G_l(\bfx,\gamma^i)=(-k)^{-l}Q_\bfk{}_{|i_1\cdots i_l}(\bfx,{\bf k}) 
  P^{i_1\cdots i_l}_{(l)}(\bfx, \gamma^i),
\nonumber
\end{equation}
and
\begin{eqnarray}
  && P_{(0)}=1,\hspace{1cm} P_{(1)}^i=\gamma^i,\hspace{1cm}
  P_{(2)}^{ij}={1\over2}(3\gamma^i\gamma^j-\gamma^{ij}),
\nonumber
\\
  &&P^{i_1\cdots i_{l+1}}_{(l+1)}
  ={2l+1\over l+1}\gamma^{(i_1}P_{(l)}^{i_2\cdots i_{l+1})}
  -{l\over l+1}\gamma^{(i_1i_2}P_{(l-1)}^{i_3\cdots i_{l+1})},
\nonumber
\end{eqnarray}
where the parentheses denote symmetrization about the indices.
For the special case of the flat universe, it reduces to
$G_l=(-i)^l \exp(i\bfx\cdot{\bf k}) P_l(\kappa^i\cdot\vecgamma)$, 
where $P_l(x)$ is the Legendre polynomial and $\kappa^i=k^i/k$.


We derive the evolution equations for the moments
$\Theta_l$ in equation (\ref{multipoledec}).
For this purpose, we use the fact that 
scalar perturbation quantities are decomposed as 
$\Phi(\eta,\bfx)=\sum_{\bf k}\Phi(\eta,k)G_0(\bfx)$, etc., 
and $\gamma_i\Vc^i=\sum_{\bf k}\Vc(\eta,k)G_1(\bfx,\gamma^i)$ 
for the velocity.
The evolution equations for the 
moments are given from equation (\ref{pertC}), by the Boltzmann hierarchy:
\begin{eqnarray}
  &&\dot\Theta_0+\xi\Theta_0=
  -{1\over 3}k\Theta_1-(\alpha+3)\dot\Phi+\xi b_X \deltac,
\label{multia}
\\
  &&\dot\Theta_1+\xi\Theta_1=
  -{2\over 5}k\Bigl(1-3{K\over k^2}\Bigr)\Theta_2+k\Theta_0
  +(\alpha+3)k\Psi+(\alpha+3)\xi b_V \Vc,
\label{multib}
\\
  &&\dot\Theta_l+\xi\Theta_l=
  -{l+1\over 2l+3}k\Bigl(1-l(l+2){K\over k^2}\Bigr)\Theta_{l+1}
  +{l\over 2l-1}k\Theta_{l-1},
\hspace{7mm}{\rm for }~(l\ge2),
\label{multic}
\end{eqnarray}
where we used 
\begin{equation}
  \gamma^i {\partial G_l\over\partial\vecx}=k\biggl(
  {l\over 2l+1}\Bigl(1-(l^2-1){K\over k^2}\Bigr) G_{l-1}
  -{l+1\over 2l+1}G_{l+1}\biggr).
\end{equation}

\section{useful formula for evolution of CDM perturbations}
\def\OmegaK{{\Omega_K}}
\def\OmegaL{{\Omega_\Lambda}}
We here summarize an useful equation for the 
evolution of the CDM perturbation (Peebles 1980). 
The CDM perturbation $\deltac(\eta,k)$ obeys 
\begin{equation}
  \ddot\deltac+{\dot a\over a}\dot \deltac
  ={3\over2}{\Omega_0H_0^2\over a}\deltac,
\end{equation}
where the expansion rate $\dot a/a$ is given by equation 
(\ref{FriedmannEq}). Note that this equation 
is independent of the scale $k$, since we are
considering the matter dominant stage.
Using equation (\ref{FriedmannEq}), we rewrite the 
above equation as
\begin{equation}
  \bigl(a\Omega_0+a^2\OmegaK+a^4\OmegaL\bigr)
  {d^2\deltac\over da^2}
  +\bigl({3\over2}\Omega_0+2a\OmegaK+3a^3\OmegaL\bigr)
  {d\deltac\over da}
  -{3\over2}{\Omega_0\over a}\deltac=0.
\end{equation}
The growing mode solution is written as 
\begin{equation}
  \deltac={5\Omega_0\over2}
  \sqrt{{\Omega_0\over a^3}+{\OmegaK\over a^2}+\OmegaL}
  \int_0^a da'
  \biggl({a'\over\Omega_0+a'\OmegaK+a'^3\OmegaL}\biggr)^{3/2}~,
\end{equation}
which is normalized as $\deltac=a$ at $a\ll1$.

\section{Analytic Approach}
\def\hatk{{\hat k}}
It is very instructive to evaluate the fluctuations in the CXB 
in an analytic way. 
To perform the analytic calculation, we consider the standard CDM model.
In the case of $\Omega_0=1$, we have $a=(\eta/\eta_0)^2$ with
$\eta_0=2/H_0$ from equation (\ref{FriedmannEq}).
Here we use the notations $\eta_{\rm max}/\eta_0=(1+z_{\rm min})^{-1/2}$ 
and $\eta_{\rm min}/\eta_0=(1+z_{\rm max})^{-1/2}$.
Omitting the terms of the source bulk motion and the gravitational 
potential fluctuations in equation (\ref{solutflat}), we have
\begin{equation}
  {\Theta_l(\eta_0,k)\over 2l+1}={b_X\over\varepsilon(\eta_0)}
  \int_{0}^{\eta_{0}}
  d\eta~\dot\varepsilon\delta_c(\eta,k)~j_{l}(k(\eta_0-\eta)).
\end{equation}
From the definition of $\varepsilon$, equation (\ref{defepsilon}),
we can write
\begin{equation}
  {\Theta_l(\eta_0,k)\over 2l+1}={b_X \delta_c(\eta_0,k)\over {\cal E}}
  \int_{\eta_{\rm min}}^{\eta_{\rm max}}d\eta
  \biggl({\eta \over \eta_0}\biggr)^{2(2+\alpha-p)} 
  j_{l}(k(\eta_0-\eta)),
\label{apptheta}
\end{equation}
with
\begin{eqnarray}
  {\cal E} &=& \int_{\eta_{\rm min}}^{\eta_{\rm max}}d\eta
  \biggl({\eta\over\eta_0}\biggr)^{2(1+\alpha-p)}\\
\nonumber
  &=& {\eta_0\over 3+2\alpha-2p}
  \biggl[(1+z_{\rm min})^{-3/2-\alpha+p}-(1+z_{\rm max})^{-3/2-\alpha+p}
  \biggr]~,
\end{eqnarray}
where we used the relation $\delta_c(\eta,k)=\delta_c(\eta_0,k) a(\eta)$.
Putting (\ref{apptheta}) into (\ref{Cl}), we have
\begin{equation}
  C_l = {\frac{2}{\pi}}{\left(b_X \over{ \cal E}\right)^2}
  \int_{0}^{\infty}dk~P(k)~
  \biggl\vert{\int_{x_1}^{x_2}
  dx~\left(1-\frac{x}{k\eta_0}\right)^{2(2+\alpha-p)}j_{l}(x)\biggr\vert}^2,
\end{equation}
where $x_1=k(\eta_0-\eta_{\rm max})$, $x_2=k(\eta_0-\eta_{\rm min})$,
and we wrote $P(k)=\bigl<{\delta_c (\eta_0,k)}^2\bigr>$.
Since $k$ integration is effective at $k\eta_0\gg1$, 
 we approximate it as
\begin{equation}
  C_l\simeq {\frac{2}{\pi}}{\left({b_X } 
  \over {\cal E} \right)^2}
  \int_{0}^{\infty}dk~P(k)~
  \biggl\vert{\int_{x_1}^{x_2}
  dx~j_{l}(x)\biggr\vert}^2.
\label{Cl2ban}
\end{equation}
This approximation works within $\sim$ a few $\times10\%$  
for $0\lesssim p\lesssim3$ and $z_{\rm min}\lesssim0.1$. 
Of course this transformation is exact when $p=2+\alpha$.

In the case of $l=1$, $x$ integration can be done. We get
\begin{equation} 
    C_{l=1}\simeq {\frac{2}{\pi}}{\left({b_X } 
  \over {\cal E} \right)^2}
  \int_{0}^{\infty}dk~P(k)~
  \biggl(j_{0}(x_1)-j_{0}(x_2)\biggr)^2.
\end{equation}
Furthermore we neglect the term of $j_{0}(x_2)$ assuming 
that $k$ integration is effective at $k\eta_0\gg1$.
We get  
\begin{equation} 
    C_{l=1}\simeq {\frac{2}{\pi}}{\left({b_X } 
  \over {\cal E} \right)^2}
  \int_{0}^{\infty}dk~P(k)~
  j_0(x_1)^2~.
\label{C14ban}
\end{equation}
Assuming $z_{\rm min}\ll 1$, we 
have $\eta_0-\eta_{\rm max}\simeq z_{\rm min}/H_0$, then
equation (\ref{C14ban}) 
can be written as
\begin{equation}
  C_{l=1}\simeq \pi b_X^2\delta_H^2\biggl({3+2\alpha-2p\over 
  1-(1+z_{\rm max})^{-3/2-\alpha+p}}\biggr)^2
  \int_0^\infty d \hatk \hatk T(\hatk)^2 
  j_0(\hatk z_{\rm min})^2~.
\label{C15ban}
\end{equation}
where $\hatk=k/H_0$, $\delta_H=1.94 \times 10^{-5}$ (Bunn \& White 1997), 
and the matter transfer function $T(k)$ is introduced.
This equation is meaningful to understand how the
dipole depends on the X-ray sources parameters. 
The dependence of the dipole on $p$ and $z_{\rm max}$ originally 
comes from the isotropic component of the X-ray background radiations 
$\propto\varepsilon(\eta_0)$,  
whereas $z_{\rm min}$ affects only the fluctuation part in
$k$ integration.

Introducing a cut off parameter $k_{\rm max}$ in $k$ integration 
instead of 
the transfer function, we approximate equation (\ref{C15ban}) as
\begin{equation}
  C_{l=1}\simeq \pi b_X^2\delta_H^2\biggl({3+2\alpha-2p\over 
  1-(1+z_{\rm max})^{-3/2-\alpha+p}}\biggr)^2
  \int_0^{\hatk_{\rm max}} d \hatk \hatk
  j_0(\hatk z_{\rm min})^2~,
\label{C16ban}
\end{equation}
where $\hatk_{\rm max}=k_{\rm max}/H_0\sim O(10^2)$.
Since $j_0(x)$ oscillates at $x\gg 1$, 
$k$ integration gives different behaviour depending on the value 
$\hatk_{\rm max} z_{\rm min}$ around unity, as follows:
\[
  \int_0^{\hatk_{\rm max}} d \hatk \hatk
  j_0(\hatk z_{\rm min})^2\simeq\left\{
    \begin{array}{ll}
       \hatk^2_{\rm max}/2 &\hspace{2cm}(\hatk_{\rm max} z_{\rm min}\ll 1),
\\
       \bigl[~{\rm ln}(2\hatk_{\rm max} z_{\rm min})
       +\gamma_{\rm E}\bigr]/2z_{\rm min}^2
       &\hspace{2cm}(\hatk_{\rm max} z_{\rm min}\gg 1),
    \end{array}
  \right.
\]
where $\gamma_{\rm E}$ is the Euler constant.
Essentially, $k_{\rm max}$ corresponds to the 
peak of the power spectrum, i.e. the scale of
galaxy large scale structure $k^{-1}_{\rm max}\sim 10 \sim 10^2 {\rm Mpc}$.
The value of $C_{l=1}$ is therefore very sensitive to 
$z_{\rm min}$ around $ O(10^{-2})$.


\label{lastpage}

\newpage
\vspace{2cm}
\begin{center}
{\large \bf FIGURE CAPTIONS}
\end{center}
\noindent
{Fig.~1---  The angular power spectrum of the fluctuations for 
various effects. The labels `$\delta$', '$V$', and '$\Psi$' denote
$\sqrt{C_l}$ due to the source clustering, 
the peculiar motion of the sources, 
and the gravitational potential fluctuations, respectively. 
Here we set $b_X=b_V=1$.
The cosmological parameters are taken as
$h=0.5$ and $\Omega_0=1$ for SCDM model, 
and $h=0.7$ and $\Omega_0=0.3$ for $\Lambda$CDM model.
The cases $z_{\rm min}=0,~0.02,~0.1$ are shown.
}

\vspace{0.5cm}
\noindent
{Fig.~2---  $\Omega_0$-dependence of $\sqrt{C}_{l=1}/b_X$
for $\Lambda$CDM and OCDM models.
The Hubble parameter $h=0.7$ is taken here.
Each panel shows the case $z_{\rm min}=0,~0.02,~0.1$, respectively.
}

\vspace{0.5cm}
\noindent
{Fig.~3---  $p$-dependence of $\sqrt{C}_{l=1}/b_X$.
The cosmological parameters are taken as
$h=0.5$ and $\Omega_0=1$ for the SCDM model,
and as $h=0.7$ and $\Omega_0=0.3$ for the $\Lambda$CDM and OCDM models.
Each panel shows the cases $z_{\rm min}=0,~0.02,~0.1$, respectively.
}

\vspace{0.5cm}
\noindent
{Fig.~4---  $z_{\rm max}$-dependence of $\sqrt{C}_{l=1}/b_X$.
The cosmological parameters are the same as Fig.~3.
}

\vspace{0.5cm}
\noindent
{Fig.~5---  $z_{\rm min}$-dependence of $\sqrt{C}_{l=1}/b_X$.
The cosmological parameters are the same as Fig.~3.
}

\end{document}